\documentclass[twocolumn,english,aps, pra, eqsecnum, longbibliography]{revtex4-2}
\usepackage[T1]{fontenc}
\usepackage[latin9]{inputenc}
\usepackage{amsmath}
\usepackage{amssymb}
\usepackage{graphicx}
\usepackage{wasysym}
\usepackage{babel}
\begin{document}
\title{Simulating Gaussian boson sampling quantum computers}
\author{Alexander S. Dellios, Margaret D. Reid and Peter D. Drummond}
\affiliation{Centre for Quantum Science and Technology Theory, Swinburne University
of Technology, Melbourne 3122, Australia}
\begin{abstract}
A growing cohort of experimental linear photonic networks implementing
Gaussian boson sampling (GBS) have now claimed quantum advantage.
However, many open questions remain on how to effectively verify these
experimental results, as scalable methods are needed that fully capture
the rich array of quantum correlations generated by these photonic
quantum computers. In this paper, we briefly review recent theoretical
methods to simulate experimental GBS networks. We focus mostly on
methods that use phase-space representations of quantum mechanics,
as these methods are highly scalable and can be used to validate experimental
outputs and claims of quantum advantage for a variety of input states,
ranging from the ideal pure squeezed vacuum state to more realistic
thermalized squeezed states. A brief overview of the theory of GBS,
recent experiments and other types of methods are also presented.
Although this is not an exhaustive review, we aim to provide a brief
introduction to phase-space methods applied to linear photonic networks
to encourage further theoretical investigations.
\end{abstract}
\maketitle

\section{Introduction}

Since Feynman's original proposal in the early 1980s \citep{Feynman_1982},
a growing amount of research has been conducted to develop large scale,
programmable quantum computers that promise to solve classically intractable
problems. Although a variety of systems have been proposed to achieve
this \citep{ciracQuantumComputationsCold1995,ciracScalableQuantumComputer2000,KnillLaflammeMilburn,devoretSuperconductingCircuitsQuantum2013a},
practical issues such as noise and losses introduce errors that scale
with system size, hampering physical implementations. Therefore, one
approach of current experimental efforts has been the development
of specialized devices aiming to unequivocally demonstrate quantum
advantage, even when experimental imperfections are present.

To this end, recent developments use photonic networks implementing
different computationally hard tasks \citep{AaronsonArkhipov:2011,Hamilton2017PhysRevLett.119.170501,quesada2018gaussian,yamamoto2017coherent}.
Such devices are made entirely from linear optics such as polarizing
beamsplitters, mirrors and phase-shifters \citep{KnillLaflammeMilburn,Kok2007},
with optical parametric oscillators (OPO) as the quantum source \citep{McMahon_CIM_science2019,Honjo2021Coherent}.
Unlike cryogenic devices based on superconducting quantum logic-gates
\citep{arute2019quantum,wuStrongQuantumComputational2021}, these
networks are operated at room temperature. Large size networks now
claim quantum advantage \citep{zhong2020quantum,zhongPhaseProgrammableGaussianBoson2021,madsenQuantumComputationalAdvantage2022,dengGaussianBosonSampling2023},
especially for a type of quantum computer called a Gaussian boson
sampler (GBS) that generates random bit outputs with a distribution
that is exponentially hard to replicate at large network sizes.

This begs the question: how can one verify that large-scale experimental
outputs are correct, when the device is designed to solve problems
no classical computer can solve in less than exponential time?

In this paper, we review theoretical methods that answer this for
photonic network based quantum computers. We especially focus on positive-P
phase-space representations that can simulate non-classical states
\citep{Drummond_generalizedP1980}. Here, we mean simulate in the
sense of a classical sampling algorithm that replicates probabilities
and moments to an accuracy better than experimental sampling errors,
for the same number of samples. Such methods account for loss and
decoherence. They are classical algorithms for standard digital computers.
In addition, they are highly scalable, with polynomial run-time for
large networks. 

Other types of simulation exist, where the classical algorithm replicates
the detection events of the experiment \citep{quesada2020exact,quesadaQuadraticSpeedUpSimulating2022,bulmerBoundaryQuantumAdvantage2022a}.
These become intractably slow for larger networks. The use of methods
like this is to offer a way to define quantum advantage, meaning a
quantum device performing tasks that are classically infeasible. Hence,
we review the definitions of simulation in the literature including
other approaches as well. There are also methods that simulate events
quickly but approximately, up to a systematic error caused by the
fact that the approximate method has a different probability distribution
\citep{villalonga2021efficient,ohSpoofingCrossEntropy2022a,oh2023tensor}.
A comparison is made of different approaches that are proposed or
used for these problems, including classical ``faking'' methods
that replicate stochastic detection events with some degree of inaccuracy.

Positive-P methods are exactly equivalent to quantum moments in the
large sample limit, and are applicable to a variety of experimental
networks. One example is the coherent Ising machine (CIM) \citep{kiesewetter2022phase,Kiesewetter2022Coherent},
which is used to solve $NP$-hard optimization problems by simulating
an Ising model with a series of light pulses \citep{wang2013coherent,yamamoto2017coherent,yamamura2017quantum}.
The largest experimental implementation of the CIM to date contains
$100,000$ modes \citep{Honjo2021Coherent}. This gives approximate
solutions to hard max-cut problems three orders of magnitude faster
than a digital computer. For such problems, typically no exact results
are known. Therefore, these solutions can be hard to validate as well.

Gaussian boson sampling (GBS) \citep{drummondSimulatingComplexNetworks2022,delliosSimulatingMacroscopicQuantum2022a,delliosValidationTestsGBS2023}
experiments have seen impressive advancements in recent years. These
specialized linear network quantum computers are used as random number
generators, where the detected output photon counts correspond to
sampling from the $\#P$-hard Hafnian or Torontonian distributions
\citep{Hamilton2017PhysRevLett.119.170501,kruse2019detailed,quesada2018gaussian},
which are either directly or indirectly related to the matrix permanent.
While more specialized than quantum circuit devices, they are also
somewhat more straightforward to implement and analyze theoretically. 

Only a few years after the original theoretical proposals \citep{Hamilton2017PhysRevLett.119.170501,quesada2018gaussian,kruse2019detailed},
multiple large scale experiments have implemented GBS with each claiming
quantum advantage \citep{zhong2020quantum,zhongPhaseProgrammableGaussianBoson2021,madsenQuantumComputationalAdvantage2022,dengGaussianBosonSampling2023}.
This rapid increase in network size has far outpaced direct verification
methods that aim to reproduce experiments using classical algorithms
\citep{quesada2020exact,quesadaQuadraticSpeedUpSimulating2022,bulmerBoundaryQuantumAdvantage2022a}.
Although other validation methods exist \citep{villalonga2021efficient,ohSpoofingCrossEntropy2022a},
these typically cannot reproduce the higher order correlations generated
in linear photonic networks of this type. 

In summary, we review progress on simulating binned correlations or
moments of experimental GBS networks using the positive-P phase-space
representation \citep{drummondSimulatingComplexNetworks2022,delliosSimulatingMacroscopicQuantum2022a,delliosValidationTestsGBS2023},
as well as comparing this with other methods. We find that the experimental
results disagree with the ideal, pure squeezed state model. However,
agreement is greatly improved with the inclusion of decoherence in
the inputs. When this decoherence is large enough, classical simulation
is feasible \citep{RahimiKeshari:2016,qi2020regimes}, and quantum
advantage is lost. Despite this, recent experimental results demonstrate
potential for quantum advantage in some of the reported datasets \citep{delliosValidationTestsGBS2023}.

In order to present a complete picture, we also review the background
theory of GBS, recent large scale experiments \citep{zhong2020quantum,zhongPhaseProgrammableGaussianBoson2021,madsenQuantumComputationalAdvantage2022,dengGaussianBosonSampling2023}
and other proposed verification algorithms \citep{bulmerBoundaryQuantumAdvantage2022a,villalonga2021efficient,quesada2020exact,quesadaQuadraticSpeedUpSimulating2022,ohSpoofingCrossEntropy2022a,oh2023tensor}. 

\section{Gaussian boson sampling: A brief overview}

The original boson sampler introduced by Aaronson and Arkhipov \citep{AaronsonArkhipov:2011}
proposed a quantum computer that generates photon count patterns by
randomly sampling an output state whose distribution corresponds to
the matrix permanent. The computational complexity arises from sampling
the permanent, which is a $\#P$-hard computational problem \citep{Aaronson2011,Scheel2004Permanents}. 

However, practical applications of the original proposal have seen
limited experimental implementations, since they require one to generate
large numbers of indistinguishable single photon Fock states. Although
small scale networks have been implemented \citep{broome2013photonic,springBosonSamplingPhotonic2013a,wangBosonSampling202019},
to reach a computationally interesting regime requires the detection
of at least 50 photons \citep{wuBenchmarkTestBoson2018,hangleiterComputationalAdvantageQuantum2022}.
This is challenging to implement experimentally. 

To solve this scalability issue, Hamilton et al \citep{Hamilton2017PhysRevLett.119.170501}
proposed replacing the Fock states with Gaussian states, which can
be more readily generated at large sizes. These types of quantum computing
networks are called Gaussian boson samplers. They still use a non-classical
input state, but one that is far easier to generate experimentally
than the original proposal of number states.

\subsection{Squeezed states}

In standard GBS, $N$ single-mode squeezed vacuum states are sent
into an $M$-mode linear photonic network. If $N\neq M$, the remaining
$N-M$ ports are vacuum states. If each input mode is independent,
the entire input state is defined via the density matrix 
\begin{equation}
\hat{\rho}^{(\text{in})}=\prod_{j}^{M}\left|r_{j}\right\rangle \left\langle r_{j}\right|.\label{eq:input_state}
\end{equation}
Here, $\left|r_{j}\right\rangle =\hat{S}(r_{j})\left|0\right\rangle $
is the squeezed vacuum state and $\hat{S}(r_{j})=\exp(r_{j}(\hat{a}_{j}^{\dagger(\text{in})})^{2}/2-r_{j}(\hat{a}_{j}^{(\text{in})})^{2}/2)$
is the squeezing operator \citep{yuen1976two,stolerEquivalenceClassesMinimum1970,stolerEquivalenceClassesMinimumUncertainty1971}.
We assume the squeezing phase is zero and $\hat{a}^{\dagger(\text{in})}$
$\left(\hat{a}_{j}^{(\text{in})}\right)$ is the input creation (annihilation)
operator for the $j$-th mode. The vacuum state $\left|0\right\rangle $
corresponds to letting $r_{j}=0$ in the squeezing operator.

The ideal GBS assumes the inputs are pure squeezed states \citep{walls1983squeezed}.
These non-classical states are characterized by their quadrature variance
and for normal ordering are defined as \citep{Kim_squeezed_PRA1989,drummond2014quantum}
\begin{align}
\Delta_{x_{j}}^{2} & =2\left(n_{j}+m_{j}\right)=e^{2r_{j}}-1\nonumber \\
\Delta_{y_{j}}^{2} & =2\left(n_{j}-m_{j}\right)=e^{-2r_{j}}-1.
\end{align}
Here, $\Delta_{x_{j}}^{2}=\left\langle (\Delta\hat{x}_{j})^{2}\right\rangle =\left\langle \hat{x}_{j}^{2}\right\rangle -\left\langle \hat{x}_{j}\right\rangle ^{2}$
is the simplified variance notation for the input quadrature operators
$\hat{x}_{j}=\hat{a}_{j}^{(\text{in})}+\hat{a}_{j}^{\dagger(\text{in})}$
and $\hat{y}_{j}=-i(\hat{a}_{j}^{(\text{in})}-\hat{a}_{j}^{\dagger(\text{in})})$,
which obey the commutation relation $[\hat{x}_{j},\hat{y}_{k}]=2i\delta_{jk}$. 

Squeezed states, while still satisfying the Heisenberg uncertainty
principle $\Delta_{x_{j}}^{2}\Delta_{y_{j}}^{2}=1$, have altered
variances such that one quadrature has variance below the vacuum noise
level, $\Delta_{x_{j}}^{2}<1$. By the Heisenberg uncertainty principle,
this causes the variance in the corresponding quadrature to become
amplified well above the vacuum limit, $\Delta_{y_{j}}^{2}>1$ \citep{yuen1976two,Drummond2004_book,vahlbruch2016detection}.
Other squeezing orientations are available and remain non-classical
so long as one variance is below the vacuum limit. 

For pure squeezed states, the number of photons in each mode is $n_{j}=\left\langle \hat{a}_{j}^{\dagger(\text{in})}\hat{a}_{j}^{(\text{in})}\right\rangle =\sinh^{2}(r_{j})$,
whilst the coherence per mode is $m_{j}=\left\langle (\hat{a}_{j}^{(\text{in})})^{2}\right\rangle =\cosh(r_{j})\sinh(r_{j})=\sqrt{n_{j}\left(n_{j}+1\right)}$.

\subsection{Input-output theory\label{subsec:Input-output-theory}}

Linear photonic networks made from a series of polarizing beamsplitters,
phase shifters and mirrors act as large scale interferometers, and
crucially conserve the Gaussian nature of the input state. 

\begin{figure}
\begin{centering}
\includegraphics[width=0.5\textwidth]{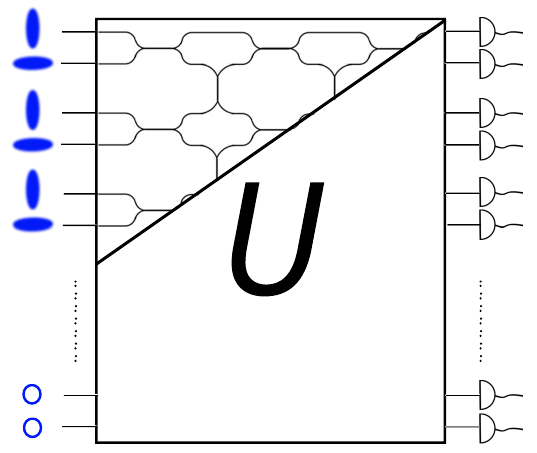}
\par\end{centering}
\caption{Gaussian boson sampling schematic where $N\subset M$ single-mode
squeezed states of different polarizations (blue ellipses) are sent
into a lossless linear photonic network where the remaining $N-M$
inputs are vacuum states (blue outlined circles). The linear network
is created from a series of phase-shifters and beamsplitters, represented
here in a circuit diagram format. This generates an exponential number
of interference pathways. Mathematically, the network is defined by
an $M\times M$ Haar random unitary $\boldsymbol{U}$. Output photon
patterns are then obtained using either PNR or threshold detectors.
\label{fig:network_schematic}}

\end{figure}

In the ideal lossless case, as seen in Fig.(\ref{fig:network_schematic}),
photonic networks are defined by an $M\times M$ Haar random unitary
matrix $\boldsymbol{U}$. The term Haar corresponds to the Haar measure,
which in general is a uniform probability measure on the elements
of a unitary matrix \citep{walschaersStatisticalBenchmarksQuantum}.

For lossless networks, input modes are converted to outputs following
\begin{equation}
\hat{a}_{i}^{(\text{out})}=\sum_{j=1}^{M}U_{ij}\hat{a}_{j}^{(\text{in})},
\end{equation}
where $\hat{a}_{i}^{(\text{out})}$ is the output annihilation operator
for the $i$-th mode. Therefore, each output mode is a linear combination
of all input modes.

Practical applications will always include some form of losses in
the network, for example photon absorption. This causes the matrix
to become non-unitary, in which case its denoted by the transmission
matrix $\boldsymbol{T}$, and the input-to-output mode conversion
now contains additional loss terms \citep{garcia-patronSimulatingBosonSampling2019,shchesnovichDistinguishingNoisyBoson2021}:
\begin{equation}
\hat{a}_{i}^{(\text{out})}=\sum_{j=1}^{N}T_{ij}\hat{a}_{j}^{(\text{in})}+\sum_{j=1}^{M}B_{ij}\hat{b}_{j}^{(\text{in})}.\label{eq:lossy_linear_transformation}
\end{equation}
Here, $\hat{b}_{j}^{(\text{in})}$ is the annihilation operator for
the $j$-th loss mode whilst $\boldsymbol{B}$ is a random noise matrix.
The loss matrix conserves the total unitary of the system as $\boldsymbol{T}\boldsymbol{T}^{\dagger}+\boldsymbol{B}\boldsymbol{B}^{\dagger}=\boldsymbol{I}$,
where $\boldsymbol{I}$ is the identity matrix. 

Although linear networks are conceptually simple systems, they introduce
computational complexity from the exponential number of possible interference
pathways photons can take in the network. These are generated from
the beamsplitters. As an example, the input ports of a 50/50 beamsplitter
are defined as superpositions of the output modes:
\begin{align}
\hat{a}_{1}^{\dagger(\text{in})} & =\frac{1}{\sqrt{2}}\left(\hat{a}_{3}^{\dagger(\text{out})}+\hat{a}_{4}^{\dagger(\text{out})}\right)\nonumber \\
\hat{a}_{2}^{\dagger(\text{in})} & =\frac{1}{\sqrt{2}}\left(\hat{a}_{3}^{\dagger(\text{out})}-\hat{a}_{4}^{\dagger(\text{out})}\right).
\end{align}

Classical states such as coherent and thermal states input into large
scale linear networks are readily simulated on a classical computer
\citep{RahimiKeshari:2016}. This leads to the question, why are non-classical
states such as Fock and squeezed states computationally hard to simulate
on large scale photonic networks? 

In short, non-classical states input into a linear network generate
large amounts of quantum correlations, which are non-trivial to simulate
classically if combined with photo-detection. Continuing with the
beamsplitter example above, if input ports $1$ and $2$ now contain
indistinguishable Fock states, as is the case with standard boson
sampling, the input state can be written as 
\begin{equation}
\left|1,1\right\rangle _{1,2}=\frac{1}{\sqrt{2}}\left(\left|2,0\right\rangle _{3,4}-\left|0,2\right\rangle _{3,4}\right).
\end{equation}
The absence of a $\left|1,1\right\rangle _{3,4}=\hat{a}_{3}^{\dagger(\text{out})}\hat{a}_{4}^{\dagger(\text{out})}\left|0,0\right\rangle _{3,4}$
term means the output photons are bunched, that is, they are entangled
and always arrive in pairs \citep{walls2008quantum}. This phenomena
is known as the Hong-Ou-Mandel (HOM) interference effect and is named
after the authors who first observed this phenomena \citep{HongPhysRevLett.59.2044}. 

If input ports $1,2$ have input squeezed vacuum states, one generates
two-mode Einstein-Podolsky-Rosen (EPR) entanglement in the quadrature
phase amplitudes \citep{Reid:1989,furusawa1998unconditional}. More
details, including a summary of the derivation, can be found in \citep{teh2021full,delliosSimulatingMacroscopicQuantum2022a}.

\subsection{Photon counting}

In GBS experiments, photon count patterns, denoted by the vector $\boldsymbol{c}$,
are generated by measuring the output state $\hat{\rho}^{(\text{out})}$
. The corresponding distribution changes depending on the type of
detector used. 

The original proposal for GBS \citep{Hamilton2017PhysRevLett.119.170501}
involved normally ordered photon-number-resolving (PNR) detectors
that can distinguish between incoming photons and implement the projection
operator \citep{walls2008quantum,Sperling2012True}
\begin{equation}
\hat{p}_{j}(c_{j})=\frac{1}{c_{j}!}:(\hat{n}'_{j})^{c_{j}}e^{-\hat{n}'_{j}}:.
\end{equation}
Here, $:\dots:$ denotes normal ordering, $c_{j}=0,1,2,\dots$ is
the number of counts in the $j$-th detector and $\hat{n}'_{j}=\hat{a}_{j}^{\dagger(\text{out})}\hat{a}_{j}^{(\text{out})}$
is the output photon number. Practically, a PNR detector typically
saturates for a maximum number of counts $c_{\text{max}}$, the value
of which varies depending on the type of PNR detector implemented. 

When the mean number of output photons per mode satisfies $\left\langle \hat{n}'_{j}\right\rangle \ll1$,
PNR detectors are equivalent to threshold, or click, detectors which
saturate for more than one photon at a detector. Threshold detectors
are described by the projection operator \citep{Sperling2012True}
\begin{equation}
\hat{\pi}_{j}(c_{j})=:e^{-\hat{n}'_{j}}\left(e^{\hat{n}'_{j}}-1\right)^{c_{j}}:,\label{eq:click_proj_op}
\end{equation}
where $c_{j}=0,1$ regardless of the actual number of photons arriving
at the detector.

In situations where $\left\langle \hat{n}'_{j}\right\rangle \gg1$,
threshold detectors become ineffective since they cannot accurate
discriminate between the different numbers of incoming photons. However,
by sending photons through a demultiplexer (demux) before arriving
at the detector, the output pulses of light become 'diluted' such
that the mean photon number is small enough to be accurately counted
using threshold detectors \citep{achillesFiberassistedDetectionPhoton2003,fitchPhotonnumberResolutionUsing2003}.
In this approach the demux acts as a type of secondary optical network
of size $1\times M_{S}$, where $M_{S}$ is the number of secondary
modes. For output photons to become sufficiently dilute, one must
satisfy $M_{S}\gg\left\langle \hat{n}'_{j}\right\rangle $ \citep{achillesFiberassistedDetectionPhoton2003,fitchPhotonnumberResolutionUsing2003,provaznikBenchmarkingPhotonNumber2020}.
When this is the case, the detectors can be thought of pseudo-PNR
(PPNR), because the threshold detectors are accurately measuring photon
counts, thus becoming equivalent to PNR detectors \citep{provaznikBenchmarkingPhotonNumber2020}. 

The computational complexity of GBS arises in the probabilities of
count patterns $\boldsymbol{c}$. These probabilities are determined
using a straightforward extension of the projection operators for
both PNR and click detectors. In the case of PNR detectors, the operator
for a specific photon count pattern is
\begin{equation}
\hat{P}(\boldsymbol{c})=\bigotimes_{j=1}^{M}\hat{p}_{j}(c_{j}),
\end{equation}
the expectation value of which corresponds to the $\#P$-hard matrix
Hafnian \citep{Hamilton2017PhysRevLett.119.170501,kruse2019detailed}
\begin{equation}
\left\langle \hat{P}(\boldsymbol{c})\right\rangle \approx|\text{Haf}(\boldsymbol{D}_{S})|^{2}.
\end{equation}
Here, $\boldsymbol{D}_{S}$ is the sub-matrix of $\boldsymbol{D}=\boldsymbol{T}\left(\bigoplus_{j=1}^{M}\tanh(r_{j})\right)\boldsymbol{T}^{T}$,
corresponding to the channels with detected photons, where the superscript
$T$ denotes the transpose of the lossy experimental transmission
matrix.

The Hafnian is a more general function than the matrix permanent,
although both are related following \citep{Hamilton2017PhysRevLett.119.170501,kruse2019detailed}
\begin{equation}
\text{perm}(\boldsymbol{D})=\text{Haf}\left(\begin{array}{cc}
0 & \boldsymbol{D}\\
\boldsymbol{D}^{T} & 0
\end{array}\right).
\end{equation}
Therefore, one of the advantages of GBS with PNR detectors is that
one can compute the permanent of large matrices from the Hafnian instead
of performing Fock state boson sampling. 

For threshold detectors, the operator for a specific binary count
pattern $\boldsymbol{c}$ is
\begin{equation}
\hat{\Pi}(\boldsymbol{c})=\bigotimes_{j=1}^{M}\hat{\pi}_{j}(c_{j}),\label{eq:Click_pattern_proj}
\end{equation}
whose expectation value corresponds to the Torontonian \citep{quesada2018gaussian}
\begin{equation}
\left\langle \hat{\Pi}(\boldsymbol{c})\right\rangle =\text{Tor}(\boldsymbol{O}_{S}),
\end{equation}
where $\boldsymbol{O}_{S}$ is a matrix constructed from the covariance
matrix of the output state. The Torontonian has also been shown to
be $\#P$-hard to compute, and is directly related to the Hafnian
\citep{quesada2018gaussian}.

\subsection{Experimental progress }

Only a few short years after the original proposal of GBS with threshold
detectors \citep{quesada2018gaussian}, the first large scale experimental
network, called Jiuzhang 1.0, was implemented by Zhong et al \citep{zhong2020quantum}.
In this experiment, $N=50$ single-mode squeezed states were sent
into a lossy $M=100$ mode linear network which generated over 50
million binary count patterns in $\approx200\text{s}$. 

The experimentalists estimated it would take Fugaku, the largest supercomputer
at the time, $600$ million years to generate the same number of samples
using the fastest exact classical algorithm at the time \citep{quesada2020exact}.
Exact samplers are algorithms designed to replicate an experiment
by directly sampling the Torontonian or Hafnian using a classical
supercomputer. When experiments claim quantum advantage, they typically
do so by comparing the run-time of the experiment with such exact
samplers. 

The algorithm implemented in \citep{quesada2020exact} uses a chain
rule sampler that samples each mode sequentially by computing the
conditional probability of observing photons in the $k$-th mode given
the photon probabilities of the previous $k-1$-th modes. The computational
complexity of this algorithm scales as $\mathcal{O}(N_{D}^{3}2^{N_{D}})$
where $N_{D}=\sum_{j=1}^{M}c_{j}$ is the total number of detected
photon counts. Due to this scaling with detected counts and system
size, more recent experiments aim to beat these exact samplers by
increasing the number of observed counts. The mean number of counts
per pattern in Jiuzhang 1.0 was $\approx41$, and the largest number
of counts in a single pattern was $76$.

Shortly after this initial experiment, in order to reduce the probability
of an exact sampler replicating the experiment, Zhong et al \citep{zhongPhaseProgrammableGaussianBoson2021}
implemented an even larger GBS network, named Jiuzhang 2.0, which
increased the number of modes to $M=144$ and performed multiple experiments
for different input laser powers and waists. This expectedly produced
an increase in the number of observed clicks, with a mean click rate
of $\approx68$ and a maximum observed click number of $N_{D}=113$.

At the same time, in classical computing, the increased probability
of multiple photons arriving at the same detector was exploited by
Bulmer et al \citep{bulmerBoundaryQuantumAdvantage2022a} to improve
the scaling of the chain rule sampler, achieving $\mathcal{O}(N_{D}^{3}2^{N_{D}/2})$
for GBS with both PNR and click detectors. Despite this apparently
modest improvement, it was estimated that generating the same number
of samples as Jiuzhang 1.0 on Fugaku would now only take $\sim73$
days \citep{bulmerBoundaryQuantumAdvantage2022a}. Due to the substantial
speed-up over previous exact algorithms \citep{quesada2020exact,quesadaQuadraticSpeedUpSimulating2022},
it was predicted that experiments using either PNR or click detectors
needed $N_{D}\gtrsim100$ to surpass exact samplers \citep{bulmerBoundaryQuantumAdvantage2022a}. 

To reduce the probability of multiple photons arriving at a single
detector, Deng et al \citep{dengGaussianBosonSampling2023} added
a $1\times8$ demux to the output modes of Jiuzhang 2.0, dubbing this
upgraded network Jiuzhang 3.0. The demux is made of multiple fiber
loop beamsplitters that separate photons into four temporal bins and
two spatial path bins and aims to ensure $\left\langle \hat{n}'_{j}\right\rangle \apprle1$,
increasing the likelihood the threshold detectors are accurately counting
the oncoming photons. This simple addition generated patterns with
almost double the largest number of clicks obtained in Jiuzhang 2.0,
with one experiment observing a maximum of $N_{D}=255$ with a mean
click number of more than $100$. 

The linear networks implemented in the three experiments above are
all static, that is, once the networks have been fabricated, one cannot
reprogram any of the internal optics. This has the advantage of circumventing
the exponential accumulation of loss that arises from increased depth
\citep{madsenQuantumComputationalAdvantage2022,deshpandeQuantumComputationalAdvantage2022a},
although one sacrifices programmability. 

To this end, Madsen et al \citep{madsenQuantumComputationalAdvantage2022}
implemented a programmable GBS called Borealis using three fiber loop
variable beamsplitters, each containing a programmable phase-shifter.
In this experiment, $N$ squeezed state pulses are generated at a
rate of $6\text{MHz}$ and are sent into each variable beamsplitter
sequentially, allowing pulses in different time-bins to interfere.
Photons output from the final fibre loop are then sent into a $1\times16$
demultiplexer and then counted by $16$ PNR detectors, which are superconducting
transition edge sensors (TES). 

The demultiplexer is required for two reasons. The first is to partially
'dilute' the number of photons in each output pulse for the PNR detectors
used \citep{madsenQuantumComputationalAdvantage2022}. The second
is to ensure each TES has reached its baseline operating temperature
before another pulse is detected, ensuring an accurate determination
of photon numbers. 

The two largest networks using this setup sent $N=216$ and $N=288$
input pulses into the network, detecting a mean photon number of $\approx125$
and $\approx143$, respectively. However, this network was more susceptible
to losses than the previous systems \citep{zhong2020quantum,zhongPhaseProgrammableGaussianBoson2021,dengGaussianBosonSampling2023}. 

\section{Simulation methods}

The rapid growth of experimental networks has spurred a parallel increase
in the number of algorithms proposed for validation and/or simulation
of these networks. Due to this, there are some inconsistencies in
the literature on the language used to define a simulator which runs
on a classical computer. Therefore, we clarify the definitions of
the various classical simulators proposed for validation used throughout
the rest of the review.

We first emphasize that a classical algorithm or classical simulator
is any program designed to run on a digital computer, be that a standard
desktop or supercomputer. The current algorithms proposed to validate
GBS can be defined as either strong or weak classical simulators \citep{pashayan2020estimation}.
There is a large literature on this topic, starting from early work
in quantum optics.

Weak classical simulators parallel the quantum computation of experimental
networks by sampling from the output probability distribution \citep{chabaudClassicalSimulationGaussian2021a}.
Algorithms that sample from the full Torontonian or Hafnian distributions
are exact samplers \citep{quesada2020exact,quesadaQuadraticSpeedUpSimulating2022,bulmerBoundaryQuantumAdvantage2022a},
as outlined above, and all known exact algorithms are exponentially
slow. There are faster but approximate algorithms producing discrete
photon count patterns that are typically called ``faked'' or ``spoofed''
patterns. Some of these generate fake patterns by sampling from the
marginal probabilities of these distributions \citep{villalonga2021efficient,ohSpoofingCrossEntropy2022a},
in which case they are called low-order marginal samplers. Such low-order
methods are only approximate, since they don't precisely simulate
higher-order correlations \citep{villalonga2021efficient,ohSpoofingCrossEntropy2022a}. 

In contrast, strong classical simulators are a type of classical algorithm
that evaluates the output probability distribution, be that the full
or marginal probabilities, of a GBS with multiplicative accuracy,
in time polynomial in the size of the quantum computation \citep{chabaudClassicalSimulationGaussian2021a,chabaudResourcesBosonicQuantum2023}. 

Phase-space simulators have a similarity to strong classical simulators,
because the samples, which are produced with a stochastic component
representing quantum fluctuations, are used to compute probabilities
or moments of the ideal distribution. Positive-P phase-space simulations
have been widely used in quantum optics for this reason \citep{Drummond_generalizedP1980,Gardiner_Book_QNoise,Drummond:2002,deuar2007correlations,DrummondOPan2014,Drummond:2016}.
While the sampling precision is not multiplicative, it is better than
the experimental precision of GBS at the same sample number, which
is sufficient for validation. By approximating the inputs as classical
states, one can also use phase-space methods to produce fake patterns
\citep{delliosValidationTestsGBS2023}, which in case they are called
a classical P-function sampler, as explained below. 

For non-classical inputs, one can generate probabilities but not count
patterns using phase-space methods: since there are no known algorithms
to generate counts efficiently from non-classical phase-space samples. 

\subsection{Exact classical simulation}

There are a number of ``exact'' classical simulation methods that
generate equivalent photon-counts, where we use quotes for ``exact'',
because even these methods have round-off errors due to to finite
arithmetic. These algorithms are all exponentially slow \citep{quesada2020exact,quesadaQuadraticSpeedUpSimulating2022,bulmerBoundaryQuantumAdvantage2022a}.
It is known that the non-classicality of the squeezed input states
of GBS is essential to providing quantum advantage in the GBS experiments,
since classical states which have a positive Glauber-Sudarshan P-representation
\citep{titulaer1965correlation,Reid1986} are known to have classically
simulable photon counts. This will be treated in Section (\ref{subsec:Classical-P-function-samplers}).
However, the non-Gaussianity arising from the photon-detection measurement
process is also important.

This is because the measurement set-up where quadrature-phase-amplitude
measurements are made on a system described by Gaussian states can
be modeled by a positive Wigner function, and hence becomes classically
simulable \citep{RahimiKeshari:2016,qi2020regimes}. Gaussian states
are defined as those with a Gaussian Wigner function. We note that
squeezed states, while non-classical, are Gaussian states. Examples
of the classical simulation of entanglement for systems with a positive
Wigner function are well known \citep{Graham1973,Steel1998,Opanchuk2014Detecting,ng2019nonlocal,lun2019phase,Opanchuk2019Mesoscopic}.

The role of the measurement set-up in GBS is clarified by the work
of Chabaud, Ferrini, Grosshans and Markham \citep{chabaudClassicalSimulationGaussian2021a}
and Chabaud and Walschaers \citep{chabaudResourcesBosonicQuantum2023}.
These authors illustrate a connection between quantum advantage and
the non-Gaussianity of both the input states and measurement set-up.
In \citep{chabaudClassicalSimulationGaussian2021a}, conditions sufficient
for the strong simulation of Gaussian quantum circuits with non-Gaussian
input states are derived. Non-Gaussian states are those for which
the Gaussian factorization of correlation functions is not applicable
\citep{lachman2022quantum}.

By demonstrating a mapping between bosonic quantum computation and
continuous-variable sampling computation, where the measurement comprises
a double quadrature detection, Chabaud and Walschaers \citep{chabaudResourcesBosonicQuantum2023}
adapt classical algorithms derived in \citep{chabaudResourcesBosonicQuantum2023}
to derive a general algorithm for the strong simulation of bosonic
quantum computation, which includes Gaussian boson sampling. They
prove that the complexity of this algorithm scales with a measure
of non-Gaussianity, the stellar rank of both the input state and measurement
set-up. This enables a quantification of non-Gaussianity, including
from the nature of the measurement, as a resource for achieving quantum
advantage in bosonic computation.

\subsection{Approximate classical simulation}

\subsubsection{Low-order marginal samplers}

Recent experiments have surpassed the threshold of applicability of
exact samplers \citep{zhongPhaseProgrammableGaussianBoson2021,madsenQuantumComputationalAdvantage2022,dengGaussianBosonSampling2023}.
Even for experiments below this threshold \citep{zhong2020quantum},
the computation time can still be very long and resource intensive.
Therefore, approximate methods that are more readily scaled to larger
size have been proposed. These cannot validate GBS, but they are useful
in quantifying a computational advantage. 

One approach presented by Villalonga et al \citep{villalonga2021efficient},
exploits the computational efficiency of computing low-order marginal
probabilities of the ideal GBS distribution to generate photon count
patterns by sampling from a distribution that contains the correct
low-order marginals. We note that the relevant marginal probabilities
must are first be evaluated before sampling. 

Two methods are implemented to compute marginals, which take the form
of multivariate cumulants, also called connected correlations or Ursell
functions. The first method corresponds to a Boltzmann machine (BM)
and computes pattern probabilities of the form \citep{villalonga2021efficient}
\begin{align}
\left\langle \hat{\Pi}(\boldsymbol{c})\right\rangle  & =\frac{1}{Z}\exp\left(\sum_{i}\lambda_{i}\hat{\pi}_{i}(c_{i})+\sum_{i<j}\lambda_{i,j}\hat{\pi}_{i}(c_{i})\hat{\pi}_{j}(c_{j})\right.\nonumber \\
 & \left.+\sum_{i<j<k}\lambda_{i,j,k}\hat{\pi}_{i}(c_{i})\hat{\pi}_{j}(c_{j})\hat{\pi}_{k}(c_{k})+\dots\right)
\end{align}
where $Z$ is the partition function that normalizes the distribution
and $\lambda_{i},\lambda_{i,j},\dots$ are parameters computed using
a mean-field approximation. 

Each term in the exponent corresponds to a marginal probability of
the ideal distribution, where a BM spoofer using only the first two
summations were implemented due to scalability. The BM method is only
applicable to GBS with threshold detectors, where faked binary patterns
are obtained via Gibbs sampling \citep{villalonga2021efficient}. 

The second method uses a greedy heuristic to generate discrete binary
patterns with approximately correct second and third-order marginals.
This algorithm scales exponentially with the desired order of the
marginal and the length of the patterns, which are typically equal
to the number of modes. Although the greedy method was originally
developed for GBS with threshold detectors, it has since been extended
to PNR detectors \citep{madsenQuantumComputationalAdvantage2022}. 

Faked patterns generated from both methods were compared to experimental
samples from Jiuzhang 1.0 and 2.0. The best comparison results came
from computing the total variational distance 

\begin{equation}
\Delta\delta=\delta_{m}-\delta_{e},
\end{equation}
where $\delta_{m}$ is the difference between pattern probabilities
of the ideal distribution and the marginal spoofers, and $\delta_{e}$
is the difference between experiment and ideal. 

Since computing pattern probabilities is computationally challenging,
comparisons were limited to only 14 modes. However, although the first-order
BM sampler was always beaten by the experiment, faked samples generated
from the second-order BM and second and third-order greedy algorithm
were closer to the ideal distribution than the experiment.

Comparisons of the cross-entropy measure, which is a widely used quantum
advantage benchmark in random circuit sampling quantum computers \citep{arute2019quantum,boixo2018characterizing,aaronsonClassicalHardnessSpoofing2020},
were also performed by Villalonga et al \citep{villalonga2021efficient}.
It remains an open question whether this is a useful measure for boson
sampling networks \citep{ohSpoofingCrossEntropy2022a}. 

In general, if experimental samples obtain a smaller cross entropy
than the spoofer, there is evidence of computational advantage. As
was the case for the total variational distance, when compared to
samples from the first-order BM, experimental samples were immediately
larger, but for all other spoofers results were mixed, with all samples
obtaining a similar cross entropy to the experiment \citep{villalonga2021efficient}. 

An algorithm introduced by Oh et al \citep{ohSpoofingCrossEntropy2022a}
was specifically designed to spoof the cross entropy measure of GBS
networks using up to second-order marginal probabilities. This algorithm
produced samples that successfully spoofed the small scale test networks
in Ref.\citep{madsenQuantumComputationalAdvantage2022}, but became
too computationally demanding to spoof larger scale networks that
claim quantum advantage \citep{zhong2020quantum,zhongPhaseProgrammableGaussianBoson2021,madsenQuantumComputationalAdvantage2022}. 

\subsubsection{Classical P-function samplers\label{subsec:Classical-P-function-samplers}}

Practical implementations of GBS will inevitably produce decoherence
and losses. These may arise within the network, as discussed above,
or in the input states as discussed in subsection \ref{subsec:GCP}.
For large enough decoherence and loss, either the inputs, outputs
or both are transformed to classical states \citep{qi2020regimes,garcia-patronSimulatingBosonSampling2019}.

For such classical states, GBS loses its quantum advantage and an
efficient classical simulation is possible. This was first shown by
Rahimi-Keshari et al \citep{RahimiKeshari:2016} for Fock state boson
sampling and later extended to GBS by Qi et al \citep{qi2020regimes},
although the fundamental ideas are much older than this \citep{Glauber1963_CoherentStates}.
In both cases, the condition for classical simulation hinges on whether
the resulting phase-space output distribution, simulated using a set
of experimental inputs, was non-negative. These conditions hinge on
the well known result that, for some states, the Glauber-Sudarshan
and Wigner representations produce negative distributions on phase-space
\citep{Hillery_Review_1984_DistributionFunctions,walls2008quantum},
which is why they are typically referred to as quasi-probability distributions. 

We define any classical algorithm that samples from the output distribution
of a classical state GBS as a classical GBS sampler. Although the
most commonly tested classical state is the fully decoherent thermal
state, it is an extreme case and and has been thoroughly disproved
for all current experimental networks \citep{zhong2020quantum,zhongPhaseProgrammableGaussianBoson2021,madsenQuantumComputationalAdvantage2022,dengGaussianBosonSampling2023}.
However, a more realistic scenario is a classical approximation to
pure squeezed states. 

\begin{figure}
\begin{centering}
\includegraphics[width=0.5\textwidth]{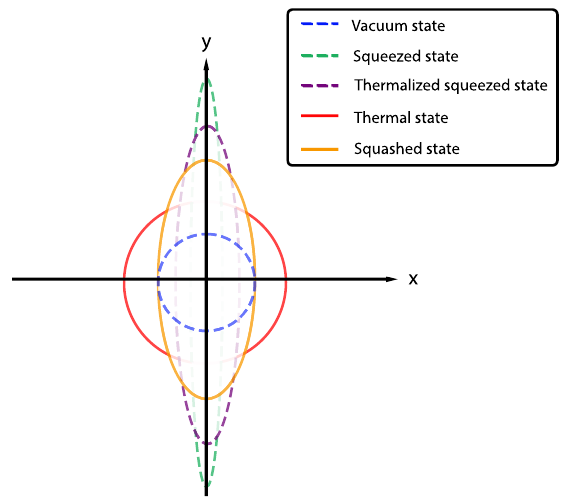}
\par\end{centering}
\caption{A diagram of the quadrature variance for different states. The vacuum
state (blue dashed line) has variance $\Delta_{x_{j}}^{2}=\Delta_{y_{j}}^{2}=1$
and arises from the Heisenberg uncertainty principle as the minimum
uncertainty state. When $\Delta_{x_{j}}^{2}<1,\:\Delta_{y_{j}}^{2}>1$
one obtains the non-classical pure squeezed state (green dashed line)
or the thermalized squeezed state (purple dashed line), which has
larger variance for the squeezed quadrature than pure squeezing but
remains below the vacuum limit. The classical thermal state (solid
red line) has variance $\Delta_{x_{j}}^{2}=\Delta_{y_{j}}^{2}>1$,
while although the squashed state has variance $\Delta_{x_{j}}^{2}\protect\neq\Delta_{y_{j}}^{2}$,
neither squeezing is below the vacuum limit as, in this case, $\Delta_{x_{j}}^{2}=1\:\Delta_{y_{j}}^{2}>1$.\label{fig:variance_comparisons}}
\end{figure}

Two such states have recently been proposed called squashed and squished
states \citep{martinez-cifuentesClassicalModelsAre2022,dengGaussianBosonSampling2023}.
Unlike thermal states, which have a quadrature variance of $\Delta_{x_{j}}^{2}=\Delta_{y_{j}}^{2}$,
squashed and squished states maintain the unequal variances of squeezed
states (see Fig.(\ref{fig:variance_comparisons}) for a diagram of
the variances for different states). Despite this, decoherence has
caused the once squeezed quadrature variance to become $\Delta_{x_{j}}^{2}=1$,
whilst $\Delta_{y_{j}}^{2}>1$. Therefore, squashed and squished states
are classical states. No true squeezing occurs, since neither variance
is below the vacuum limit. We note that the difference between these
two states is that squished states must contain the same mean number
of photons as the input squeezed state \citep{dengGaussianBosonSampling2023},
whereas the squashed state photon number can vary. 

Comparisons of all experimental networks with simulated thermal states
have, expectedly, failed \citep{zhong2020quantum,zhongPhaseProgrammableGaussianBoson2021,madsenQuantumComputationalAdvantage2022,dengGaussianBosonSampling2023}.
However, Juizhang 1.0 data was shown to have a large degree of input
decoherence \citep{drummondSimulatingComplexNetworks2022}, and hence
simulated squashed states were shown to model samples from Jiuzhang
1.0 as well as the theoretical ideal distribution for some statistical
tests \citep{martinez-cifuentesClassicalModelsAre2022}. An efficient
phase-space sampler, which can generate binary patterns for any classical
input state, later showed that squashed state count patterns were
closer to the simulated ideal distribution than the experimental data
set in Jiuzhang 2.0 that claims quantum advantage \citep{delliosValidationTestsGBS2023}.
This method is reviewed in more detail in section \ref{sec:phase-space GBS}.
The most recent GBS experiments with PNR and click detectors, Borealis
and Jiuzhang 3.0, produced outputs that were closer to the ideal than
simulated squashed and squished states, although squished states have
only been tested against samples from Jiuzhang 3.0 to date. 

The classical GBS samplers implemented in these tests assume either
all $N\subset M$ inputs or all $M$ outputs are classical states.
A more physically realistic scenario is one where the inputs and/or
outputs are mixtures of classical and quantum states. In order to
model this, a classical sampler that scales with the loss rate $\eta$
of the linear network was introduced by Oh et al \citep{oh2023tensor}. 

The aim of this algorithm is to simulate the output covariance matrix
$\boldsymbol{V}_{p}$ of a pure squeezed state GBS using a tensor
network method. This covariance matrix arises from a decomposition
of the actual output covariance matrix $\boldsymbol{V}=\boldsymbol{W}+\boldsymbol{V}_{p}$,
where $\boldsymbol{W}\geq0$ is a random displacement matrix arising
from classical photons in the outputs. 

As outlined in the previous subsection, the computational complexity
of computing the ideal GBS, corresponding to $\boldsymbol{V}_{p}$
in this case, scales with the number of system modes and hence detected
photons. However, only non-classical output photons contribute to
this computational complexity. Therefore, if more experimental output
photons are classical rather than non-classical, the matrix $\boldsymbol{W}$
will dominate the computation of $\boldsymbol{V}$ \citep{oh2023tensor},
which one should then be able to simulate efficiently. 

This is the general principle of the algorithm implemented by Oh et
al \citep{oh2023tensor}, where the number of non-classical output
photons are computed from the actual squeezing parameter $s=-1/2\ln(1-\eta)$.
Clearly, as the loss rate increases, the non-classical photons decrease,
in turn causing the number of classical photons to increase, allowing
$\boldsymbol{W}$ to dominate the output covariance matrix. 

Samples obtained from this method were used to calculate the total
variational distance and cross entropy for the small scale test networks
of Ref. \citep{madsenQuantumComputationalAdvantage2022}, as well
as second and higher-order cumulants for larger scale networks \citep{zhong2020quantum,zhongPhaseProgrammableGaussianBoson2021,madsenQuantumComputationalAdvantage2022,dengGaussianBosonSampling2023}.
In all cases, samples produced from this classical sampler were closer
to the ideal distributions than all the experiments, highlighting
the extent to which the loss rate $\eta$ play a role in affecting
claims of quantum advantage in the current generation of GBS. 

\section{Validating Gaussian boson sampling in phase-space\label{sec:phase-space GBS}}

One of the many open questions in GBS research is how to efficiently
validate large scale experimental networks. The exact methods reviewed
above either suffer from scalability issues \citep{quesada2020exact,quesadaQuadraticSpeedUpSimulating2022,bulmerBoundaryQuantumAdvantage2022a},
or else they don't simulate higher-order correlations \citep{villalonga2021efficient}
due to computational complexity. Due to the rapid growth of experimental
networks, even higher order correlations will be produced due to the
increase in interference pathways. The requirement of a simulation
that validates a quantum computer is that it allows the computation
of measurable probabilities with an accuracy at least equal to the
experimental sampling error.

Such correlations and probabilities play an increasingly important
role in characterization of the output distribution \citep{Moran},
even when losses are present \citep{shchesnovichDistinguishingNoisyBoson2021,shchesnovichBosonSamplingCannot2022},
despite their increased sensitivity to decoherence. Therefore, scalable
methods are needed that simulate higher-order correlations without
performing the impractical $\#P$-hard direct computation of the count
samples. To this end, we review recent theoretical methods that simulate
grouped count probabilities (GCPs) of GBS networks using the normally
ordered positive-P phase-space representation. 

These methods can be used for networks with threshold detectors, and
have successfully been used to compare theory and experiment for samples
from Jiuzhang 1.0 \citep{drummondSimulatingComplexNetworks2022} and
Jiuzhang 2.0 \citep{delliosValidationTestsGBS2023}. We emphasize
that the positive-P representation does not produce discrete count
patterns. However, the simulated output distributions have identical
moments to the ideal GBS distributions. 

Before we briefly review previous results presented in \citep{drummondSimulatingComplexNetworks2022},
we first outline the necessary background theory on phase-space representations,
focusing on normally ordered methods, and GCPs. The interested reader
is referred to Refs. \citep{drummondSimulatingComplexNetworks2022,delliosSimulatingMacroscopicQuantum2022a,delliosValidationTestsGBS2023}
for more details, while the highly efficient phase-space code, xQSim,
can be downloaded from public domain repositories \citep{GitHubPeterddrummondXqsim}. 

\subsection{Phase-space methods}

Originally developed by Wigner for symmetrically ordered operators
\citep{Wigner_1932}, phase-space representations establish a mapping
between quantum operators of different orderings to probability distributions
defined on the classical phase-space variables of position and momentum
\citep{Carmichael_Book1,walls2008quantum}, which are more commonly
rewritten in terms of the complex coherent state amplitude vectors
$\boldsymbol{\alpha}$, $\boldsymbol{\alpha}^{*}$. 

Moyal \citep{Moyal_1949} later showed how one can use these methods
can be used to compute the dynamics of operators. Due to this, phase-space
representations are frequently used to compute the operator master
equation \citep{gardiner2004quantum,Carmichael_Book1}, which for
some representations, corresponds to the second-order Fokker-Planck
equation (FPE), which is commonly used in statistical mechanics. 

The FPE in turn, can be mapped to a stochastic differential equation
(SDE) that introduces randomly fluctuating terms and, for some real
or complex vector $\boldsymbol{a}$, takes the general form \citep{gardiner2004handbook}
\begin{equation}
\frac{d}{dt}\boldsymbol{a}=\boldsymbol{A}(\boldsymbol{a})+\boldsymbol{B}(\boldsymbol{a})\xi(t),\label{eq:SDE}
\end{equation}
where $\boldsymbol{A}$ is a vector function and $\boldsymbol{B}$
is a matrix, both of which are typically known, while $\xi(t)$ is
a real Gaussian noise term with $\left\langle \xi\right\rangle =0$
and $\left\langle \xi_{i}(t)\xi_{j}(t')\right\rangle =\delta(t-t')\delta_{ij}$.
These randomly fluctuating terms, defined as the derivative of a Wiener
process \citep{gardiner2004handbook}, play an analogous role to quantum
and thermal fluctuations for applications in quantum mechanics. 

Each solution of an SDE is a stochastic trajectories in phase-space
and physically corresponds to a single experiment. Therefore, averages
over an ensemble of trajectories $E_{S}$ corresponds to a mean value
from multiple experimental shots. 

Such dynamical methods have been successfully applied to a variety
of quantum optics systems \citep{DrummondGardinerWalls1981,RosalesZarateBellPhysRevA.90.022109,teh2020overcoming,teh2020dynamics}
(also see Ref.(\citep{Drummond:2016}) for a brief review), including
the CIM \citep{kiesewetter2022phase,Kiesewetter2022Coherent}. However,
as is clear from subsection \ref{subsec:Input-output-theory}, linear
networks are not modeled dynamically and hence cannot produce an SDE. 

Instead, phase-space representations model linear networks as stochastic
difference equations (SDFE) where the $M$-th order SDFE takes the
general form \citep{rodkinaStochasticDifferenceEquations2010}
\begin{equation}
\boldsymbol{a}_{n+1}=\sum_{m=0}^{M}\boldsymbol{A}_{m}(n-m,\boldsymbol{a}_{n-m})+\boldsymbol{B}(n,\boldsymbol{a}_{n})\xi_{n},\label{eq:SDFE}
\end{equation}
where $\boldsymbol{a}_{n}$, $\boldsymbol{A}_{m}$, $\boldsymbol{B}$
and $\xi_{n}$are discrete analogs to their continues variable definitions
in Eq.(\ref{eq:SDE}). This becomes clearer when comparing with Eq.(\ref{eq:lossy_linear_transformation}). 

Due to the randomly fluctuating term, SDEs don't have analytical solutions
and must be computed numerically using a variety of methods \citep{kloedenStochasticDifferentialEquations1992}.
As is also the case with numerical computations of non-stochastic
differential equations, SDEs are approximated as difference equations
for numerical computation. Although one usually has practical issues
such small step size limits for SDEs \citep{kloedenStochasticDifferentialEquations1992,rodkinaStochasticDifferenceEquations2010},
SDEs and SDFEs are two sides of the same coin. 

Due to the use of PNR and threshold detectors, we focus on simulating
linear networks using the normally ordered Glauber-Sudarshan diagonal
P-representation and generalized P-representations. Non-normally ordered
Wigner and Q-function methods are possible also \citep{drummondSimulatingComplexNetworks2022,delliosSimulatingMacroscopicQuantum2022a,delliosValidationTestsGBS2023},
but these have too large a sampling error to be useful for simulations
of photon counting.

\subsubsection{Generalized P-representation}

The generalized P-representation developed by Drummond and Gardiner
\citep{Drummond_Gardiner_PositivePRep} is a family of normally ordered
phase-space representations that produce exact and strictly positive
distributions on phase-space for any quantum state. It was developed
as a non-classical extension of the Glauber-Sudarshan diagonal P-representation,
which is defined as 
\[
\hat{\rho}=\int P(\boldsymbol{\alpha})\left|\boldsymbol{\alpha}\right\rangle \left\langle \boldsymbol{\alpha}\right|\text{d}^{2}\boldsymbol{\alpha},
\]
where $\left|\boldsymbol{\alpha}\right\rangle $ is a coherent state
vector. 

Due to the absence of off-diagonal terms in the density matrix, which
represent quantum superpositions, the diagonal P-representation produces
a distribution $P(\boldsymbol{\alpha})$ that is negative and singular
for non-classical states such as Fock and squeezed states. 

To account for this, the generalized P-representation introduces the
projection operator 
\begin{equation}
\hat{\Lambda}(\boldsymbol{\alpha},\boldsymbol{\beta})=\frac{\left|\boldsymbol{\alpha}\right\rangle \left\langle \boldsymbol{\beta}^{*}\right|}{\left\langle \boldsymbol{\beta}^{*}|\boldsymbol{\alpha}\right\rangle },
\end{equation}
which doubles the phase-space dimension. This increased dimension
allows quantum superpositions to exist, since the basis now generates
independent coherent state amplitudes $\boldsymbol{\alpha}$, $\boldsymbol{\beta}$
with off-diagonal amplitudes $\boldsymbol{\beta}\neq\boldsymbol{\alpha}^{*}$,
which define a quantum phase-space of larger dimensionality. 

The most useful generalized-P method for simulating linear networks
with squeezed state inputs is the positive P-representation, which
expands the density matrix as the $4M$-dimensional volume integral
\begin{equation}
\hat{\rho}=\int\int P(\boldsymbol{\alpha},\boldsymbol{\beta})\hat{\Lambda}(\boldsymbol{\alpha},\boldsymbol{\beta})\text{d}^{2M}\boldsymbol{\alpha}\text{d}^{2M}\boldsymbol{\beta}.\label{eq:+P_general}
\end{equation}
Here, $\boldsymbol{\alpha}$, $\boldsymbol{\beta}$ can vary along
the entire complex plane and by taking the real part of Eq.(\ref{eq:+P_general}),
the density matrix becomes hermitian, thus allowing efficient sampling. 

The other generalized-P method, called the complex P-representation,
requires $P(\boldsymbol{\alpha},\boldsymbol{\beta})$ to be complex,
resulting from its definition as a contour integral \citep{Drummond_Gardiner_PositivePRep}.
This makes the complex P-representation useful for simulating Fock
state boson sampling, which requires a complex weight term $\Omega$
to be applied to the sampled distribution \citep{opanchuk2018simulating,drummond2020initial}. 

One of the key reasons the positive P-representation is useful for
simulating photonic networks arises from the moment relationship
\begin{align}
\left\langle \hat{a}_{j_{1}}^{\dagger}\dots\hat{a}_{j_{n}}\right\rangle  & =\left\langle \beta_{j_{1}}\dots\alpha_{j_{n}}\right\rangle _{P},\nonumber \\
 & =\int\int P(\boldsymbol{\alpha},\boldsymbol{\beta})\left(\beta_{j_{1}}\dots\alpha_{j_{n}}\right)\text{d}^{2M}\boldsymbol{\alpha}\text{d}^{2M}\boldsymbol{\beta},\label{eq:stochastic_equivalance}
\end{align}
where $\left\langle \dots\right\rangle $ denotes a quantum expectation
value and $\left\langle \dots\right\rangle _{P}$ a positive-P ensemble
average. Therefore, normally ordered operator moments are exactly
equivalent to positive-P phase-space moments, which is also valid
for any generalized P-representation. 

The reader familiar with phase-space methods may know that other representations,
such as the Wigner and Husimi Q function, also output a positive,
non-singular distribution for Gaussian, non-classical states. 

While this is certainly true, for experiments using normally ordered
detectors, one must re-order every non-normally ordered operator to
obtain normal ordering. This introduces a term corresponding to vacuum
noise in the initial phase-space samples, resulting in sampling errors
that increase exponentially for higher-order correlations, thereby
making such methods unsuitable for simulating photonic networks \citep{delliosValidationTestsGBS2023}.

Using the coherent state expansion of pure squeezed states \citep{adam1994complete},
one can define the input state Eq.(\ref{eq:input_state}) in terms
of the positive P-representation as 
\begin{equation}
\hat{\rho}^{(\text{in})}=\text{Re}\int\int P(\boldsymbol{\alpha},\boldsymbol{\beta})\hat{\Lambda}(\boldsymbol{\alpha},\boldsymbol{\beta})\text{d}\boldsymbol{\alpha}\text{d}\boldsymbol{\beta}.
\end{equation}
The resulting positive-P distribution for input pure squeezed states
is \citep{drummondSimulatingComplexNetworks2022}
\begin{equation}
P(\boldsymbol{\alpha},\boldsymbol{\beta})=\prod_{j}C_{j}e^{-(\alpha_{j}^{2}+\beta_{j}^{2})(\gamma_{j}^{-1}+1/2)+\alpha_{j}\beta_{j}},\label{eq:pure_squeezed_dist}
\end{equation}
where $C_{j}=\sqrt{1+\gamma_{j}}/(\pi\gamma_{j})$ is a normalization
constant and $\gamma_{j}=e^{2r_{j}}-1$. 

Output samples are then readily obtained by transforming the input
coherent amplitudes as $\boldsymbol{\alpha}'=\boldsymbol{T}\boldsymbol{\alpha}$,
$\boldsymbol{\beta}'=\boldsymbol{T}^{*}\boldsymbol{\beta}$, which
corresponds to sampling from the output density matrix 
\begin{equation}
\hat{\rho}^{(\text{out})}=\text{Re}\int\int P(\boldsymbol{\alpha},\boldsymbol{\beta})\hat{\Lambda}(\boldsymbol{\alpha}',\boldsymbol{\beta}')\text{d}\boldsymbol{\alpha}\text{d}\boldsymbol{\beta}.
\end{equation}

\subsection{Grouped count probabilities\label{subsec:GCP}}

For GBS with threshold detectors, the number of possible binary patterns
obtained from an experiment is $\approx2^{M}$ with each pattern having
a probability of roughly $\left\langle \hat{\Pi}(\boldsymbol{c})\right\rangle \approx2^{-M}$.
Samples from large scale networks are exponentially sparse, requiring
binning to obtain meaningful statistics. It is therefore necessary
to define a grouped count observable that corresponds to an experimentally
measurable quantity.

The most general observable of interest is a $d$-dimensional grouped
count probability (GCP), defined as \citep{drummondSimulatingComplexNetworks2022}
\begin{equation}
\mathcal{G}_{\boldsymbol{S}}^{(n)}(\boldsymbol{m})=\left\langle \prod_{j=1}^{d}\left[\sum_{\sum c_{i}=m_{j}}\hat{\Pi}_{S_{j}}(\boldsymbol{c})\right]\right\rangle ,
\end{equation}
which is the probability of observing $\boldsymbol{m}=(m_{1},\dots,m_{d})$
grouped counts of output modes $\boldsymbol{M}=(M_{1},M_{2},\dots)$
contained within disjoint subsets $\boldsymbol{S}=(S_{1},S_{2},\dots)$.
Here, similar to Glauber's original definition from intensity correlations
\citep{Glauber1963_CoherentStates}, $n=\sum_{j=1}^{d}M_{j}\leq M$
is the GCP correlation order.

Each grouped count $m_{j}$ is obtained by summing over binary patterns
$\boldsymbol{c}$ for modes contained within a subset $S_{j}$. For
example, a $d=1$ dimensional GCP, typically called total counts,
generates grouped counts as $m_{j}=\sum_{i}^{M}c_{i}$, whilst $d>1$
gives $\boldsymbol{m}=(m_{1}=\sum_{i=1}^{M/d}c_{i},\dots,m_{d}=\sum_{i=M/d+1}^{M}c_{i})$.
This definition also includes more traditional low-order marginal
count probabilities and moments.

Although a variety of observables can be generated using GCPs, such
as the marginal probabilities commonly used by spoofing algorithms
\citep{villalonga2021efficient,ohSpoofingCrossEntropy2022a}, multi-dimensional
GCPs are particularly useful for comparisons with experiment. The
increased dimension causes the number of grouped counts to grow, e.g.
for $d=2$ $\boldsymbol{m}=(m_{1},m_{2})$, which in turn increases
the number of count bins (data points) available for comparisons,
providing a fine grained comparison of results. This also causes effects
of higher order correlations to become more statistically significant
in the data \citep{delliosValidationTestsGBS2023}. 

One the most useful applications arises by randomly permuting the
modes within each subset $S_{j}$, which results in a different grouped
count for each permutation. The number of possible permutation tests
scales as \citep{delliosValidationTestsGBS2023}
\begin{equation}
\frac{\binom{M}{M/d}}{d}=\frac{M!}{d(M/d)!(M-M/d)!},
\end{equation}
where different correlations are compared in each comparison test.
This leads to a very large number of distinct tests, making good use
of the available experimental data. 

Despite these advantages, caution must be taken when comparing very
high dimensional GCPs, since the increased number of bins means that
there are fewer counts per bin. This causes the experimental sampling
errors to increase, reducing the significance of statistical tests
\citep{delliosValidationTestsGBS2023}. 

\section{Computational phase-space methods}

To simulate GCPs in phase-space, input stochastic amplitudes $\boldsymbol{\alpha}$,
$\boldsymbol{\beta}$ corresponding to the phase-space distribution
of the input state $\hat{\rho}^{(\text{in})}$ must first be generated.
However, although pure squeezed states are the theoretical 'gold standard'
of GBS, practically they are challenging to create in a laboratory
and inevitably decoherence will arise from equipment. 

\subsection{Decoherent input states}

A more realistic model is one that includes decoherence in the input
state. Such a model was proposed in \citep{drummondSimulatingComplexNetworks2022}
and assumes the intensity of the input light is weakened by a factor
of $1-\epsilon$ whilst adding $n_{j}^{\text{th}}=\epsilon n_{j}$
thermal photons per mode. The number of input photons per mode remains
unchanged from the pure squeezed state definition, but the coherence
of photons is now altered as $\tilde{m}=(1-\epsilon)m_{j}$. To account
for possible measurement errors, a correction factor $t$ is also
applied to the transmission matrix, to improve the fit to the simulated
distribution. 

The advantage of this decoherence model, which is a type of thermal
squeezed state \citep{Fearn_JModOpt1988,Kim_squeezed_PRA1989}, is
that it allows a simple method for generating phase-space samples
for any Gaussian state following 
\begin{align}
\alpha_{j} & =\frac{1}{2}\left(\Delta_{x_{j}}w_{j}+i\Delta_{y_{j}}w_{j+M}\right)\nonumber \\
\beta_{j} & =\frac{1}{2}\left(\Delta_{x_{j}}w_{j}-i\Delta_{y_{j}}w_{j+M}\right),\label{eq:input_amplitudes}
\end{align}
where $\left\langle w_{j}w_{k}\right\rangle =\delta_{jk}$ are real
Gaussian noises that model quantum fluctuations in each quadrature
and 
\begin{align}
\Delta_{x_{j}}^{2} & =2(n_{j}+\tilde{m}_{j})\nonumber \\
\Delta_{y_{j}}^{2} & =2(n_{j}-\tilde{m}_{j}),
\end{align}
are the thermal squeezed state quadrature variances. 

As $\epsilon\rightarrow1$ the inputs become classical, since $\Delta_{x_{j}}^{2},\,\Delta_{y_{j}}^{2}\geq1$,
but small amounts of thermalization causes the inputs to remain non-classical
given, for example, with a squeezing orientation of $\Delta_{x_{j}}^{2}<0$.
So long as the state is Gaussian, one can model a variety of inputs,
both classical and non-classical, by simply varying $\epsilon$, where
$\epsilon=0$ corresponds to a pure squeezed state and $\epsilon=1$
to a thermal state. 

The input stochastic samples are also straightforwardly extended to
non-normally ordered representations, as outlined in detail in \citep{drummondSimulatingComplexNetworks2022,delliosSimulatingMacroscopicQuantum2022a,delliosValidationTestsGBS2023}.

Performing the summation over binary patterns can now be efficiently
simulated in phase-space using the multi-dimensional inverse discrete
Fourier transform \citep{drummondSimulatingComplexNetworks2022}
\begin{equation}
\mathcal{G}_{\boldsymbol{S}}^{(n)}\left(\boldsymbol{m}\right)=\frac{1}{\prod_{j}(M_{j}+1)}\sum_{\boldsymbol{k}}\tilde{\mathcal{G}}_{M}^{(n)}(\boldsymbol{k})e^{i\sum_{j}k_{j}\theta_{j}m_{j}},
\end{equation}
where the Fourier observable simulates all possible correlations generated
in an experimental network and is defined as 
\begin{equation}
\mathcal{\tilde{G}}_{\boldsymbol{S}}^{(n)}\left(\boldsymbol{k}\right)=\left\langle \prod_{j=1}^{d}\bigotimes_{i\in S_{j}}\left(\pi_{i}(0)+\pi_{i}(1)e^{-ik_{j}\theta_{j}}\right)\right\rangle _{P}.
\end{equation}
Here, $\theta_{j}=2\pi/(M_{j}+1)$, $k_{j}=0,\dots,M_{j}$ and $\pi_{j}=\exp(-n'_{j})(\exp(n'_{j})-1)^{c_{j}}$
is the positive-P click observable, obtained from the equivalence
Eq.(\ref{eq:stochastic_equivalance}), where $n'_{j}=\alpha'_{j}\beta'_{j}$
is the output photon number. 

This simulation method is highly scalable, with most observables taking
a few minutes to simulate on a desktop computer for current experimental
scales. To highlight this scalability, much larger simulations of
network sizes of up to $M=16,000$ modes have also been performed
\citep{drummondSimulatingComplexNetworks2022}. 

\subsection{Phase-space classical sampler}

If classical states are input into a GBS, the network can be efficiently
simulated using the diagonal P-representation, which arises as a special
case of the positive P-representation if $P(\boldsymbol{\alpha},\boldsymbol{\beta})=P(\boldsymbol{\alpha})\delta(\boldsymbol{\alpha}^{*}-\boldsymbol{\beta})$.
Due to this, initial stochastic samples are still generated using
Eq.(\ref{eq:input_amplitudes}), except now one has rotated to a classical
phase-space with $\boldsymbol{\beta}=\boldsymbol{\alpha}^{*}$. 

Similar to non-classical states, the input density matrix for any
classical state is defined as 
\begin{equation}
\hat{\rho}^{(\text{in})}=\int P(\boldsymbol{\alpha})\left|\boldsymbol{\alpha}\right\rangle \left\langle \boldsymbol{\alpha}\right|\text{d}^{2}\boldsymbol{\alpha},
\end{equation}
where the form of the distribution $P(\boldsymbol{\alpha})$ changes
depending on the state. Input amplitudes are again transformed to
outputs following $\boldsymbol{\alpha}'=\boldsymbol{T}\boldsymbol{\alpha}$
and are used to define the output state
\begin{equation}
\hat{\rho}^{(\text{out})}=\int P(\boldsymbol{\alpha})\left|\boldsymbol{\alpha}'\right\rangle \left\langle \boldsymbol{\alpha}'\right|\text{d}^{2}\boldsymbol{\alpha}.\label{eq:diag_P_density}
\end{equation}

Using the output coherent amplitudes, one can now efficiently generate
binary patterns corresponding to any classical state input into a
linear network. In order to conserve the simulated counts for each
ensemble, corresponding to a single experimental shot, the $j$-th
output mode of the $k$-th ensemble is independently and randomly
sampled via the Bernoulli distribution \citep{delliosValidationTestsGBS2023}
\begin{equation}
P_{j}^{(k)}(c_{j}^{(\text{class)}})=(p_{j}^{(k)})^{c_{j}^{(\text{class)}}}(1-p_{j}^{(k)})^{1-c_{j}^{(\text{class)}}}.
\end{equation}
Here, $c_{j}^{(\text{class})}=0,1$ is the classically generated bit
of the $j$-th mode where the probability of $c_{j}^{(\text{class})}=1$,
the 'success' probability, is 
\begin{equation}
p_{j}^{(k)}=(\pi_{j}(1))^{(k)}=\left(1-e^{-n'_{j}}\right)^{(k)}.
\end{equation}
This is simply the click probability of the $k$-th stochastic ensemble
with an output photon number of $n'_{j}=\left|\alpha_{j}\right|^{2}$. 

For an $M$-mode network, each stochastic ensemble outputs the classical
faked pattern of the form 
\begin{equation}
(\boldsymbol{c}^{(\text{class})})^{(k)}=[P_{1}^{(k)},P_{2}^{(k)},\dots,P_{M}^{(k)}],
\end{equation}
which are binned to obtain GCPs, denoted $\mathcal{G}^{(\text{class})}$,
that can be compared to simulated distributions. 

\begin{figure}
\begin{centering}
\includegraphics[width=0.5\textwidth]{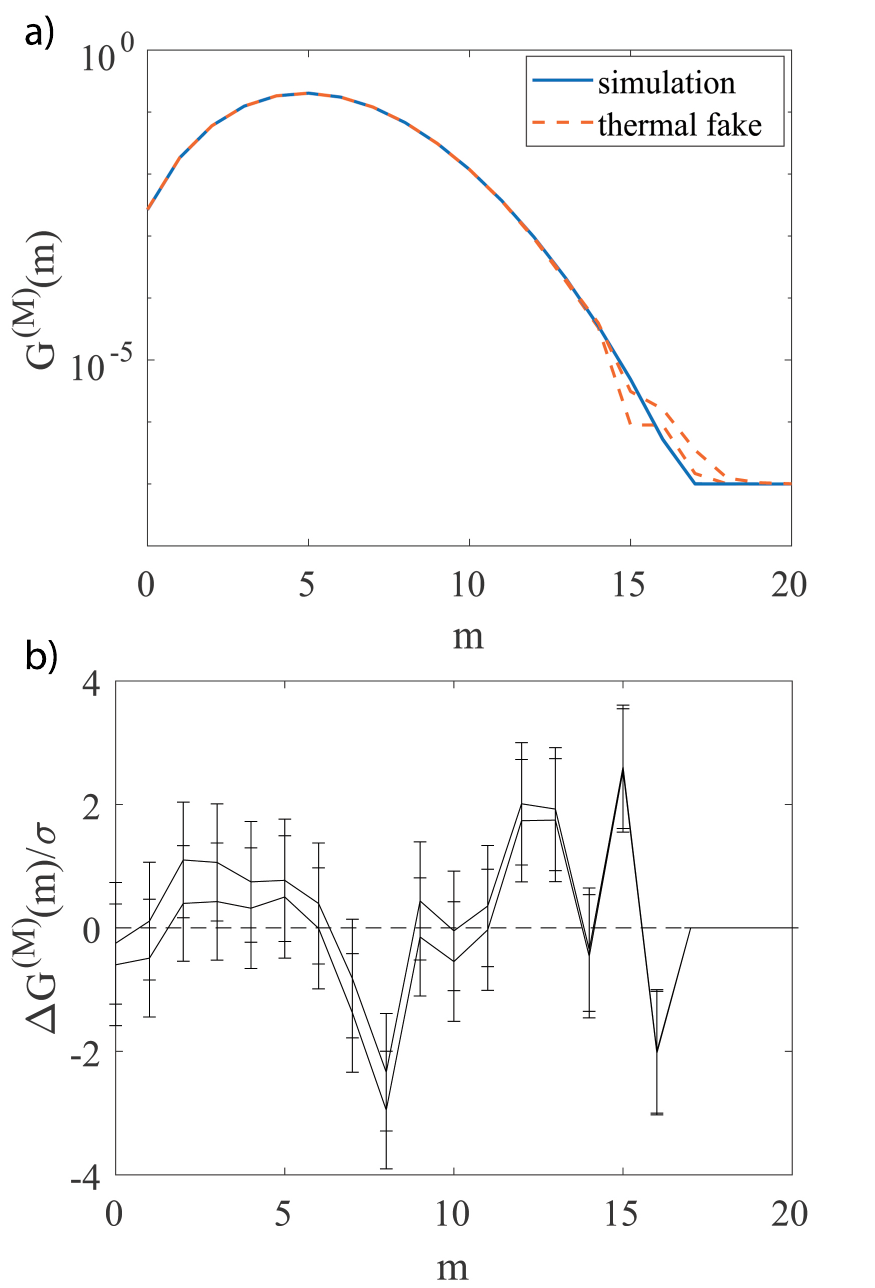}
\par\end{centering}
\caption{Comparisons of total counts, the $m$ count probability regardless
of pattern, for $4\times10^{6}$ classical binary patterns generated
from thermal states with $\boldsymbol{r}=[1,\dots,1]$ input into
a $20\times20$ Haar random unitary matrix $\boldsymbol{U}$ with
transmission coefficient $t=0.5$. a) Positive-P phase-space simulated
output distribution, obtained using $E_{S}=1.2\times10^{6}$ ensembles,
for thermal states sent into a linear network denoted by the solid
blue line are compared against the distribution produced by the binned
thermal fake patterns which are represented by the orange dashed line.
b) Comparison of the normalized difference $\Delta\mathcal{G}^{(M)}(\boldsymbol{m})/\sigma_{m}=\left(\bar{\mathcal{G}}-\mathcal{G}^{\text{(class)}}\right)/\sigma_{m}$,
where $\sigma_{m}$ is the sum over compared grouped count variances
(following from Eq.(\ref{eq:chi-square})), and $\mathcal{G}^{\text{(class)}}$
represents a classically simulated, discrete count probability. Upper
and lower lines correspond to one standard deviation of theoretical
phase-space sampling errors, while the error-bars correspond to the
sampling errors of the simulated ``fake'' counts. The results are
cut off for photon count bins containing less than 10 counts. The
expected good agreement becomes clearer when Z-score statistical tests
are performed, giving $Z^{(\text{class})}\approx1$. In this example,
since the input is classical there is no quantum advantage, and the
classical sampler is exact, as expected. \label{fig:Comparisons-of-total-thermal}}

\end{figure}

In the limit of large numbers of patterns, binned classical fakes
are approximately equal to the simulated output distribution corresponding
to the density matrix Eq. (\ref{eq:diag_P_density}). An example of
this can be seen in Fig. (\ref{fig:Comparisons-of-total-thermal})
where $4\times10^{6}$ binary patterns are generated by sending $N=20$
thermal states into an $M=N$ mode lossless linear network represented
by a Haar random unitary. The binned classical patterns produce a
total count distribution that agrees with simulations within sampling
error, because for this classical input example there is no quantum
advantage.

\subsection{Statistical tests}

For comparisons to be meaningful, statistical tests are needed to
quantify differences between experiment and theory. To this end, Refs.
\citep{drummondSimulatingComplexNetworks2022,delliosSimulatingMacroscopicQuantum2022a}
implement a slightly altered version of a standard chi-square test,
which is frequently used in random number generator validation \citep{Rukhin2010,knuth2014art},
and is defined as 
\begin{equation}
\chi_{ES}^{2}=\sum_{i=1}^{k}\frac{\left(\mathcal{G}_{i,E}-\bar{\mathcal{G}}_{i,S}\right)^{2}}{\sigma_{i}^{2}}.\label{eq:chi-square}
\end{equation}
Here, $k$ is the number of valid photon count bins, which we define
as a bin with more than $10$ counts, and $\mathcal{\bar{G}}_{i,S}$
is the phase-space simulated GCP ensemble mean for any input state
$S$, which converges to the true theoretical GCP, $\mathcal{G}_{i,S}$,
in the limit $E_{S}\rightarrow\infty$. The experimental GCP is defined
as $\mathcal{G}_{i,E}$, and $\sigma_{i}^{2}=\sigma_{T,i}^{2}+\sigma_{E,i}^{2}$
is the sum of experimental, $\sigma_{E,i}^{2}$, and theoretical,
$\sigma_{T,i}^{2}$, sampling errors of the $i$-th bin. 

The chi-square test result follows a chi-square distribution that
converges to a normal distribution when $k\rightarrow\infty$ due
to the central limit theorem \citep{wilsonDistributionChiSquare1931,johnsonContinuousUnivariateDistributions1970}.
One can use this to introduce another test that determines how many
standard deviations away the result is from its mean. The aim of this
test is to obtain the probability of observing the output $\chi_{ES}^{2}$
result using standard probability theory. For example, an output of
$6\sigma$, where $\sigma$ is the standard deviation of the normal
distribution, indicates the data has a very small probability of being
observed. 

To do this, Ref.\citep{delliosValidationTestsGBS2023} implemented
the Z-score, or Z-statistic, test which is defined as 
\begin{equation}
Z_{ES}=\frac{\left(\chi_{ES}^{2}/k\right)^{1/3}-\left(1-2/(9k)\right)}{\sqrt{2/(9k)}},\label{eq:WH_Zstat}
\end{equation}
where $\left(\chi_{ES}^{2}/k\right)^{1/3}$ is the Wilson-Hilferty
(WH) transformed chi-square statistic \citep{wilsonDistributionChiSquare1931},
which allows a faster convergence to a normal distribution when $k\geq10$
\citep{wilsonDistributionChiSquare1931,johnsonContinuousUnivariateDistributions1970},
and $\mu=1-\sigma^{2}$, $\sigma^{2}=2/(9k)$ are the corresponding
mean and variance of the normal distribution. 

The Z-statistic allows one to determine the probability of the count
patterns. A result of $Z_{ES}>6$ would indicate that experimental
distributions are so far from the simulated results that patterns
may be displaying non-random behavior. Due to this, and to present
a unified comparison notation, we compute the Z-statistic of the chi-square
results presented in \citep{drummondSimulatingComplexNetworks2022}
using Eq.(\ref{eq:WH_Zstat}). 

Valid experimental results with correct distributions should have
$\chi_{EI}^{2}/k\approx1$, where the subscript $I$ denotes the ideal
GBS distribution, with $Z_{EI}\approx1$. For claims of quantum advantage,
one must simultaneously prove that $Z_{CI}\gg1$, where the subscript
$C$ denotes binary patterns from the best classical fake, such as
the diagonal-P method described above. This is the 'gold standard'
and would show that, within sampling errors, experimental samples
are valid, and closer to the ideal distribution than any classical
fake. 

In the more realistic scenario of thermalized squeezed inputs, one
may still have quantum advantage if $Z_{ET}\approx1$ while $Z_{CT}\gg1$.
Therefore, these four observables are of the most interest for comparisons
of theory and experiment, and are given below. 

\subsection{Comparisons with experiment}

Throughout this section, we primarily review comparisons from Jiuzhang
1.0, for purposes of illustration \citep{drummondSimulatingComplexNetworks2022}.
A thorough comparison of all data sets obtained in Jiuzhang 2.0 is
presented elsewhere \citep{delliosValidationTestsGBS2023}. 

We first review comparisons of total counts, which is the probability
of observing $m$ clicks in any pattern and is usually one of the
first observables experimentalists compare samples to. This is because
in the limit of a large number of clicks, one can estimate the ideal
distribution as a Gaussian distribution via the central limit theorem. 

To simulate the ideal total counts distribution in phase-space using
GCPs we let $n=M$ and $\boldsymbol{S}=\{1,2,\dots,M\}$, giving $\mathcal{G}_{\{1,\dots,M\}}^{(M)}(m)$.
For Jiuzhang 1.0, one obtains a Z-statistic of $Z_{EI}\approx340$
for $k=63$ valid bins. Clearly, experimental samples are far from
the ideal and indicate photon count distributions are not what would
be expected from an ideal GBS.

\begin{figure}
\begin{centering}
\includegraphics[width=0.5\textwidth]{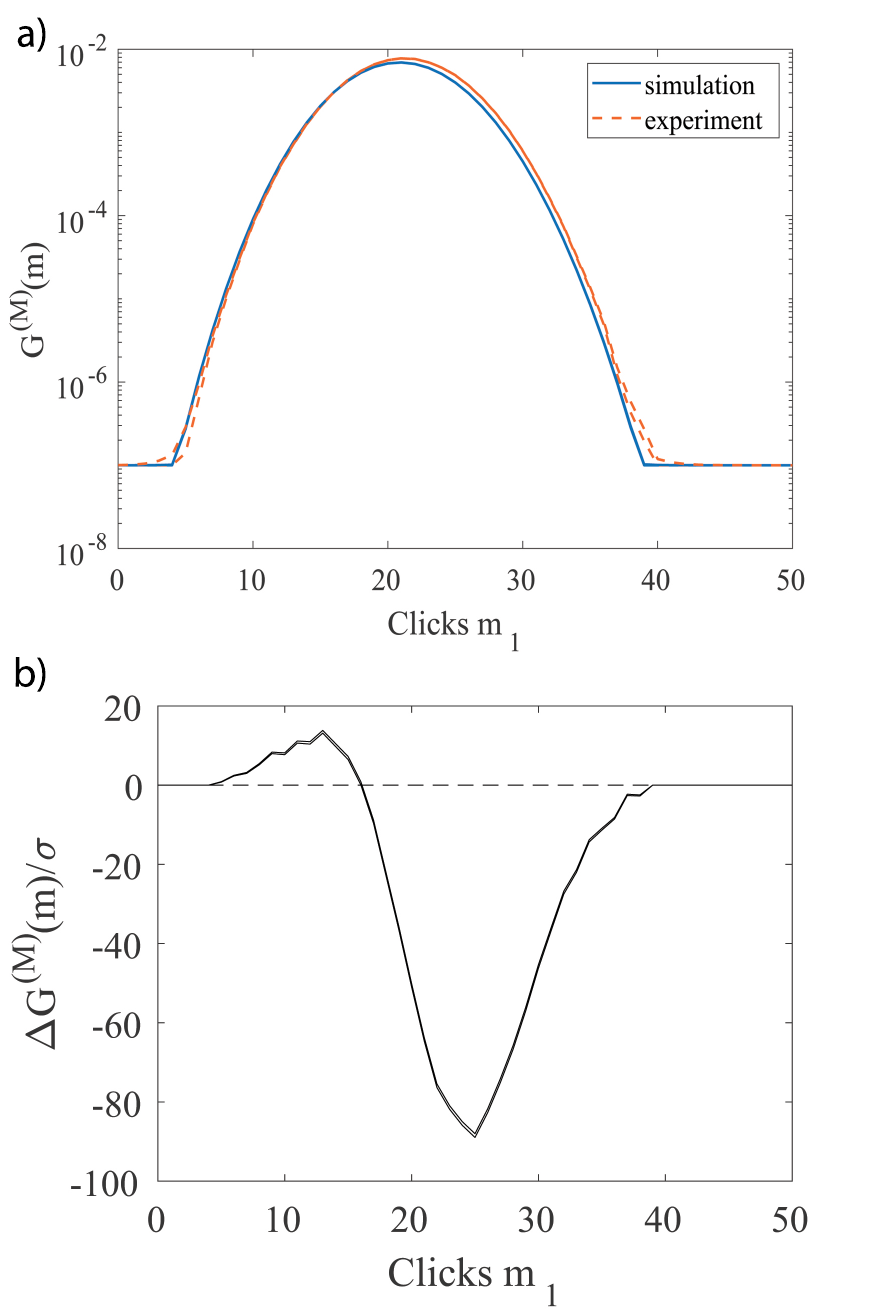}
\par\end{centering}
\caption{a) Comparison of a one-dimensional planar slice through the maximum
of a $d=2$ dimensional GCP $\mathcal{G}_{50,50}^{(100)}(m)$ for
binary patterns from \emph{Jiuzhang 1.0} data (dashed orange line).
Phase-space simulations (solid blue line) are performed for the ideal
GBS, \emph{without }decoherence, and use $E_{S}=1.2\times10^{6}$
stochastic ensembles. b) Normalized difference between theory and
experiment $\Delta\mathcal{G}^{(M)}(\boldsymbol{m})/\sigma_{m}$ versus
$m_{1}$, which is the grouped count for modes contained within the
first subset $S_{1}$. Upper and lower solid black lines are $\pm1\sigma_{T}$,
with grouped count bins containing less than $10$ counts being cut
off. Although the distributions are visually similar in a), the normalized
difference shows that significant discrepancies are present (see Fig.(\ref{fig:Comparisons-of-total-thermal})
for definitions), and one can readily determine that the ideal GBS
model is not validated for this set of experimental data, although
the agreement is much better if a decoherent, thermalised model is
used instead. \label{fig:2D binning}}
\end{figure}

To determine whether these differences either increase or decrease
when higher-order correlations become more prevalent in the simulations,
the dimension of the GCP is increased to $d=2$. In this case, Jiuzhang
1.0 sees an almost doubling in Z values for comparisons with the simulated
ideal distribution (see Fig.(\ref{fig:2D binning})), where $Z_{EI}\approx647$
is obtained from $k=978$. \citep{drummondSimulatingComplexNetworks2022}.
The increase in the number of valid bins with GCP dimension causes
the normal distribution of the WH transformed chi-square statistic
to have a smaller variance. When compared to a single dimension, where
$k=63$ gives $\sigma^{2}\approx3.5\times10^{-3}$, binning with $d>1$
causes experimental samples to now pass a more stringent test, as
$k=978$ produces a normal distribution with variance $\sigma^{2}\approx2.3\times10^{-4}$. 

Simulating the more realistic scenario of thermalized squeezed states,
a thermalized component of $\epsilon=0.0932\pm0.0005$ and a transmission
matrix correction of $t=1.0235\pm0.0005$ is used as to compare a
simulated model to samples from Jiuzhang 1.0 \citep{drummondSimulatingComplexNetworks2022}.
In this case, an order of magnitude improvement in the resulting Z
value is observed where $Z_{ET}\approx17\pm2$. 

Despite this significant improvement, differences are still large
enough that $Z_{ET}>1$. Similar results were also obtained for data
sets with the largest recorded photon counts in Jiuzhang 2.0, that
is data sets claiming quantum advantage \citep{delliosValidationTestsGBS2023}.
However, this is not the case for data sets with small numbers of
recorded photons which typically give $Z_{ET}\approx1$ \citep{delliosValidationTestsGBS2023},
although these experiments should be easily replicated by exact samplers.
The large amount of apparent thermalization is the likely reason why
simulated squashed states described Jiuzhang 1.0 samples just as well
as the ideal GBS in Ref. \citep{martinez-cifuentesClassicalModelsAre2022}. 

When higher-order correlations are considered, samples from Jiuzhang
1.0 are far from the expected correlation moments of the ideal distribution.
Although including the above fitting parameters improves this result
with $Z_{ET}\approx31$, the samples still deviate significantly from
the theoretical thermalized distribution \citep{drummondSimulatingComplexNetworks2022}. 

Unlike comparisons of total counts, samples from all data sets in
Jiuzhang 2.0 satisfy $Z_{EI},Z_{ET}>1$ for simulations with $d>1$,
meaning higher-order correlations also display significant departures
from the expected theoretical results, even for the simpler cases
with low numbers of photon counts \citep{delliosValidationTestsGBS2023}.
The reasons for this are not currently known.

\section{Summary}

In order to effectively validate GBS quantum computers, scalable methods
are needed which capture the entire interference complexity of linear
networks. The positive-P phase-space representation is the only currently
known method which can simulate every measurable grouped output distribution
with non-classical input states, allowing efficient comparisons of
theory and experiment for data that is available on a reasonable time-scale. 

One of the important issues here is the extraordinary relative sparseness
of the experimental data, which makes it completely impossible to
experimentally recover any reasonable estimate of the full distribution.
Thus, while the full distribution is exponentially hard to compute,
it is equally hard to measure. This means that comparisons of theory
and experiment always involve some type of grouping or binning of
the available data.

The next significant point is that one can both experimentally measure
and theoretically compute the grouped distributions. This can be carried
out theoretically with great precision using the positive-P phase-space
simulation method, combined with a Fourier transform binning algorithm.
These do not add greatly to the computational overhead, giving exact
tests that are computable in just a few minutes, which is of great
practical value.

The resulting statistical tests employed in these comparisons are
far more scalable than ones implemented in many previous comparisons
\citep{zhong2020quantum,zhongPhaseProgrammableGaussianBoson2021,madsenQuantumComputationalAdvantage2022,dengGaussianBosonSampling2023,villalonga2021efficient,ohSpoofingCrossEntropy2022a,oh2023tensor},
as they don't require the computation of photon count pattern probabilities,
which limits the comparisons that can be performed using these tests.
Exact simulation using direct photon counts is impractical in the
large-scale regime where quantum advantage is found.

Statistical testing shows that the GBS experiments tested are far
from the ideal GBS output distributions which are obtained from injecting
pure squeezed vacuum state inputs into a linear network. The reason
for the discrepancy is that some form of decoherence is inevitable
in such large quantum systems, and this makes the ideal standard too
high a bar that is unlikely to ever be fully realized. A more reasonable
goal is an output distribution obtained from including some decoherence
in the inputs, although the amount of decoherence must be small enough
that input states remain non-classical. 

In summary, the positive P-representation provides an effective, scalable
method to validate quantum photonic network data. It is not limited
to quantum computing applications such as the GBS, as the theory presented
here can be readily adapted to other optical networks, and can include
non-ideal features that occur in practical experiments. Having a target
distribution which is non-ideal yet non-classical is the ``Goldilocks''
target calculation of these devices: a porridge which is neither too
hot nor too cold. 

\section*{Declarations}

\subsection*{Availability of data and materials}

Phase-space GBS simulation software xQSim can be downloaded from public
domain repositories at \citep{GitHubPeterddrummondXqsim}.

\subsection*{Competing interests}

The Authors declare no competing interests. 

\subsection*{Funding}

This research was funded through grants from NTT Phi Laboratories
and the Australian Research Council Discovery Program.

\subsection*{Authors' contributions}

A.S.D wrote the original draft, created figures and performed numerical
simulations. All authors contributed to reviewing and editing this
manuscript. 

\bibliographystyle{apsrev4-2}

\end{document}